\documentclass[sigconf,nonacm]{acmart}

\setcopyright{acmlicensed} %% CCS: DO NOT REMOVE
\copyrightyear{2018} %% CCS: DO NOT REMOVE
\acmYear{2018} %% CCS: DO NOT REMOVE
\acmDOI{XXXXXXX.XXXXXXX} %% CCS: DO NOT REMOVE
\acmConference[Conference acronym 'XX]{Make sure to enter the correct
  conference title from your rights confirmation email}{June 03--05,
  2018}{Woodstock, NY}  %% CCS: DO NOT REMOVE
\acmISBN{978-1-4503-XXXX-X/2018/06}  %% CCS: DO NOT REMOVE

\usepackage{booktabs} % For formal tables
\usepackage{graphicx}
\usepackage{epstopdf}
\usepackage{makecell}
\usepackage[flushleft]{threeparttable}
\usepackage{listings}
\usepackage{xcolor}
\usepackage{balance}
\usepackage{amsmath}

\usepackage{tabu}
\usepackage{xspace}
\usepackage{multirow}
\usepackage{wasysym}
\usepackage{xcolor}

\colorlet{json_key}{magenta}
\colorlet{json_value}{blue}
\colorlet{json_num}{orange}
\colorlet{json_punct}{red!60!black}

\lstdefinelanguage{json}{
    basicstyle=\ttfamily\scriptsize\color{black}, 
    showstringspaces=false,
    breaklines=true,
    breakatwhitespace=true,
    alsoletter={", :, _}, 
    morekeywords={
        "sig":, "calls":, "cfg":,
        "kotlin_PermissionService":, "kotlin_implementation":,
        "BB_0":, "BB_1":, "BB_2":, "BB_3":, "BB_4":, "BB_5":, "BB_6":, "BB_7":, "BB_8":, "BB_9":,
        "BB_10":, "BB_11":, "BB_12":, "BB_13":, "BB_14":, "BB_15":, "BB_16":, "BB_17":, "BB_18":, "BB_19":,
        "BB_20":, "BB_21":, "BB_22":, "BB_23":, "BB_24":, "BB_25":, "BB_26":, "BB_27":, "BB_28":, "BB_29":,
        "BB_30":, "BB_31":, "BB_32":, "BB_33":, "BB_34":, "BB_35":, "BB_36":, "BB_37":, "BB_38":, "BB_39":,
        "BB_40":, "BB_41":, "BB_42":, "BB_43":, "BB_44":, "BB_45":, "BB_46":, "BB_47":, "BB_48":, "BB_49":, "BB_50":,
        "divergence_id":, "analyzed_java_entry_method":, "analyzed_kotlin_entry_method":,
        "java_path_involved":, "kotlin_path_involved":, "neutral_divergence_description":,
        "is_vulnerable":, "affected_implementation":, "vulnerability_type":, 
        "severity_rating":, "severity_reasoning":, "exploit_rationale":
    },
    keywordstyle=\color{magenta}\bfseries,
    literate=
      {\{}{{{\color{black}{\{}}}}{1}
      {\}}{{{\color{black}{\}}}}}{1}
      {[}{{{\color{black}{[}}}}{1}
      {]}{{{\color{black}{]}}}}{1}
      {,}{{{\color{black}{,}}}}{1}
}

\usepackage{nimbusmononarrow}
\usepackage[T1]{fontenc}

\usepackage{textcomp,booktabs}
\usepackage{colortbl}
\usepackage{marvosym}
\usepackage{algorithm}
\usepackage{algorithmicx}
\usepackage{algpseudocode}

\usepackage{pifont}% http://ctan.org/pkg/pifont
\usepackage{listings}
\usepackage{cleveref}
\usepackage{enumitem}
\usepackage{microtype}
\usepackage{ulem}
\usepackage{fancyhdr}
\usepackage{balance}
\usepackage{tabularray} 
\usepackage{seqsplit}
\usepackage{xspace}
\usepackage{fontawesome5}
\usepackage{marvosym}
\usepackage{multirow}
\usepackage{color,xcolor,colortbl}
\usepackage{soul}
\usepackage{framed}

\usepackage{hyperref}
\hypersetup{
    colorlinks = true,
    urlcolor = black,
    linkcolor = black,
    citecolor = black
}

\usepackage[all]{nowidow}
\usepackage{xurl} % 导入xurl包
\usepackage{hyperref} % 提供高级URL处理功能

\makeatletter
\newcommand{\thickhline}{%
	\noalign {\ifnum 0=`}\fi \hrule height 0.8pt
	\futurelet \reserved@a \@xhline
}
\newcolumntype{"}{@{\hskip\tabcolsep\vrule width 0.8pt\hskip\tabcolsep}}
\newcolumntype{*}{!{\vrule width 0.8pt}}
\makeatother

\newcommand{\squishlist}{
\begin{itemize}[noitemsep,nolistsep,leftmargin=\parindent]
  \setlength{\itemsep}{-0pt}
  \setlength{\parskip}{0pt}
}
\newcommand{\squishend}{
  \end{itemize}
}
\usepackage{tikz}
\usepackage{amsmath}

\usepackage{filecontents}

\usepackage{float}

\usepackage{textcomp,booktabs}
\usepackage{colortbl}
\usepackage[all]{nowidow}

\newcommand{\paragraphNew}[1]{\vspace{3pt}\noindent{\bf{#1}.}}

\newcommand{\miniparagraph}{\vspace{2pt}\noindent}

\newcommand{\toolname}{{\sc ParaDroid}\xspace}

\usepackage{xurl}

\definecolor{mygreen}{rgb}{0,0.6,0}
\definecolor{mymauve}{rgb}{0.58,0,0.82}
\definecolor{ashgrey}{rgb}{0.7, 0.75, 0.71}
\definecolor{mygrey}{rgb}{0.85, 0.85, 0.85}

\definecolor{codegreen}{rgb}{0,0.6,0}
\definecolor{codegray}{rgb}{0.5,0.5,0.5}
\definecolor{codepurple}{rgb}{0.58,0,0.82}
\definecolor{backcolour}{rgb}{0.95,0.95,0.95}

\definecolor{mygray}{gray}{.9}
\definecolor{verbgray}{gray}{0.92}
\definecolor{shadecolor}{rgb}{.9, .9, .9}
\definecolor{brandeisblue}{rgb}{0.0, 0.44, 1.0}
\definecolor{electricpurple}{rgb}{0.75, 0.0, 1.0}

\definecolor{mygreen}{rgb}{0,0.6,0}
\definecolor{mymauve}{rgb}{0.58,0,0.82}
\definecolor{ashgrey}{rgb}{0.7, 0.75, 0.71}
\definecolor{mygrey}{rgb}{0.85, 0.85, 0.85}
\definecolor{mypink}{HTML}{FFCCCC}
\definecolor{lightred}{HTML}{FF9999}
\definecolor{mediatered}{HTML}{FF7675}
\definecolor{rqpink}{HTML}{fd79a8}
\definecolor{codegreen}{rgb}{0,0.6,0}
\definecolor{codegray}{rgb}{0.6,0.6,0.6}
\definecolor{codepurple}{rgb}{0.58,0,0.82}
\definecolor{backcolor}{HTML}{F7F7F7}
\definecolor{lightblue}{HTML}{ADD8E6}
\definecolor{backblue}{HTML}{E2EDF6}
\definecolor{backred}{HTML}{F7CBAC}
\definecolor{backorange}{HTML}{FFBD82}
\definecolor{backblue}{HTML}{78C1F3}
\definecolor{backdarkblue}{HTML}{2D74B5}
\definecolor{mygray}{gray}{.9}
\definecolor{newcolor}{HTML}{60a3bc}
\definecolor{myred}{HTML}{FF7675}
\definecolor{mylightred1}{HTML}{FAB1A0}
\definecolor{mylightred2}{HTML}{FFCCCC}

\lstdefinestyle{mystyle}{
	backgroundcolor=\color{backcolour},
	commentstyle=\color{codegreen},
	keywordstyle=\color{magenta},
	numberstyle=\ttfamily\footnotesize\color{codegray},
	stringstyle=\color{codepurple},
	mathescape=true,
	xleftmargin=8pt,
	xrightmargin=3pt,
	basicstyle=\ttfamily\small,
	breakatwhitespace=false,
	breaklines=true,
	captionpos=b,
	numbers=left,
	numbersep=5pt,
	showspaces=false,
	showstringspaces=false,
	showtabs=false,
	tabsize=2,
	frame=single,
	moredelim=**[is][\color{red}]{@}{@},
}

\lstnewenvironment{logcode}{%
	\lstset{backgroundcolor=\color{verbgray},
		frame=single,
		framerule=0pt,
		basicstyle=\ttfamily,
		columns=fullflexible}}{}

\usepackage{tcolorbox}

\tcbuselibrary{breakable, skins}

\usepackage{caption}

\tcbset{%
	label begin/.style={label={#1}},%          just for symmetry
	label end/.style={after upper=\label{#1}}% new end label
}

\newtcolorbox[%
auto counter]{mybox}[2][]{%
	enhanced jigsaw,
	breakable,
	#1}

\usepackage{wrapfig}

\begin{document}
\title[Lost in Migration]{Lost in Migration: Exposing Android Framework Vulnerabilities in Parallel Java-Kotlin Implementations}

\author{Rui Li}
\affiliation{%
  \institution{Singapore Management University}
  \country{Singapore}
}
\email{ruili@smu.edu.sg}

\author{Wenrui Diao}
\affiliation{%
  \institution{Shandong University}
  \city{Qingdao}
  \country{China}
}
\email{diaowenrui@link.cuhk.edu.hk}

\author{Debin Gao}
\affiliation{%
  \institution{Singapore Management University}
  \country{Singapore}
}
\email{dbgao@smu.edu.sg}

\ccsdesc[500]{Security and privacy~Software and application security}

\keywords{Android OS, Language Migration, Vulnerability Discovery} %% CCS: DO NOT REMOVE but you MAY update

\begin{abstract}

Android has adopted Kotlin alongside Java across apps and core system components. During this shift, we observe \textit{parallel implementations} in the Android Open Source Project (AOSP) where the same component is implemented in both Java and Kotlin. In principle, their functional purposes are identical. In practice, subtle semantic divergences can appear. Such divergences are not vulnerabilities by themselves, but they provide useful clues that may reveal flaws in surrounding enforcement logic. To the best of our knowledge, this paper presents the first systematic study of Java-Kotlin parallel implementations in the Android framework and examines their security implications. We design and build \toolname, an analysis framework that identifies parallel methods at scale and compares their behaviors. \toolname normalizes code into a bytecode-level intermediate representation, reconstructs class-to-source mappings, and uses large language models to reason about method semantics and identify behavioral divergences. Evaluated on AOSP Android 14-16, \toolname identified 329 parallel method pairs and 37 vulnerable divergences. We responsibly disclosed the exploitable issues to the Android Security Team. Three vulnerabilities and two bugs have been confirmed, and two CVE IDs have been assigned. Our results demonstrate that parallel Java-Kotlin code paths provide a practical surface for discovering security flaws in modern Android.

\end{abstract}

\maketitle

\section{Introduction}
\label{sec:introduction}

Android powers billions of devices and remains the most widely deployed mobile operating system. As the platform evolves, Google has promoted Kotlin to first-class status for Android development~\cite{url_kotlin_first}. Kotlin compiles to JVM bytecode and interoperates with existing Java code~\cite{url_kotlin}, which has encouraged its adoption across both applications and system components. The Android Open Source Project (AOSP) reflects this trend, with a growing share of Kotlin code emerging alongside the still-dominant Java codebase.

Prior work~\cite{DBLP:journals/tse/MartinezM22} and official guidance~\cite{url_kotlin} highlight Kotlin’s advantages over Java, including concise syntax, null-safety, higher-order functions, and coroutines. These features reduce boilerplate and improve maintainability, while integrating seamlessly with Java. While existing studies primarily discuss the attractiveness of Kotlin, they fail to investigate how coexisting Java and Kotlin implementations of the same subsystem may evolve differently or pose security concerns.

In recent versions of AOSP, we observe a recurring phenomenon that we term \textbf{\textit{parallel implementations}}: two codepaths, one in Java and one in Kotlin, that are designed to serve an identical functional purpose. Specifically, \textit{they aim to achieve the same operational outcome, even though their internal implementation logic and coding patterns may differ}. These pairs may reside in the same or different packages, use similar class names, or be activated at runtime through version checks. Parallel implementations often appear during staged migrations or refactoring. Conceptually, they are meant to be interchangeable. In practice, however, subtle semantic differences can emerge. \textbf{\textit{While such differences are not always vulnerabilities in isolation, they can provide critical clues that reveal flaws in the surrounding enforcement logic.}} Our motivating case (Section~\ref{subsec:motivation-case}) in the Android permission subsystem demonstrates that the divergence between the Java and Kotlin paths exposes an inconsistency in enforcement logic, which could be exploited to obtain unauthorized sensitive information.

\paragraphNew{Our Work} To the best of our knowledge, this is the first systematic analysis specifically targeting the security implications of the co-evolutionary phase in Android's Java-to-Kotlin migration. We design and build \toolname, an analysis framework that identifies parallel methods at scale and compares their behaviors. \toolname normalizes Java and Kotlin into a common bytecode-level representation to reconstruct class-to-source mappings and detect candidate parallel pairs. It then standardizes the source code of these targeted pairs into language-agnostic Unified Execution Graphs (UEGs). Finally, it leverages Large Language Models (LLMs) to reason about code-level semantics across the UEGs, identifying behavioral divergences and evaluating potential security risks.

We evaluated \toolname on AOSP versions 14, 15, and 16. The tool identified 329 parallel method pairs. Within this set, it detected 372 behavioral divergences and flagged 37 cases as potentially security-relevant. Following deduplication and manual review, we identified 11 unique, exploitable vulnerabilities and reported them to the Android Security Team. At the time of submission, Google confirmed five distinct issues. They classified three findings as security vulnerabilities. Two of these vulnerabilities received high-severity ratings and were assigned CVE IDs. One vulnerability received a low-severity rating. Google confirmed the two other issues as functional defects.

\paragraphNew{Contributions} The main contributions of this work are:

\begin{itemize}[leftmargin = 15pt, topsep=2pt]
	
    \item \textit{Parallel implementation phenomenon.} This work exposes parallel Java-Kotlin implementations in AOSP as a systematic source of semantic divergences. These discrepancies serve as critical indicators for uncovering security flaws in surrounding enforcement logic.

	\item \textit{Analysis framework.} We designed and implemented \toolname to automatically locate parallel method pairs. It compares their behaviors using bytecode normalization, class-source recovery, a language-agnostic unified execution graph, and LLM-assisted semantic reasoning.
	
    \item \textit{Real-world vulnerabilities.} This work uncovered security-relevant divergences in core framework components. Attackers can exploit these flaws to achieve privilege escalation and data leakage. Google acknowledged these findings and assigned CVE IDs.

\end{itemize}

\paragraphNew{Roadmap} Section~\ref{sec:background} provides background on Kotlin for Android and the threat model. Section~\ref{sec:security} presents a motivating case and outlines automated analysis challenges. Section~\ref{sec:design} details the design and implementation of \toolname. Section~\ref{sec:results} presents the evaluation results. Section~\ref{sec:bugs} analyzes discovered vulnerabilities through case studies. Section~\ref{sec:discussion} covers limitations and mitigation strategies. Sections~\ref{sec:relatedwork} and \ref{sec:conclusion} review related work and conclude the paper.

\section{Background and Threat Model}
\label{sec:background}

This section provides the necessary background on the Kotlin language and its role in the Android ecosystem. We also introduce the threat model assumed in this paper.

\subsection{Kotlin for Android}
\label{subsec:kotlin}

Kotlin~\cite{url_kotlin} is a modern statically typed programming language introduced by JetBrains in 2011. It compiles to Java bytecode and interoperates seamlessly with Java. In 2017, Google announced official support for Kotlin in Android, and in 2019, launched the “Kotlin First” strategy~\cite{url_kotlin_first} to promote its adoption in both Android apps and system development.

Compared to Java, Kotlin introduces features designed to improve code quality and developer experience. Its concise syntax reduces boilerplate, such as verbose getters, setters, and explicit type declarations~\cite{url_Kotlin_VS_Java_Difference}. Kotlin also enforces strict null-safety within its type system, helping prevent common errors like null pointer exceptions~\cite{url_kotlin_first}. Additionally, Kotlin supports lambda expressions and extension functions, facilitating functional programming patterns~\cite{url_Kotlin_lambdas}. It further provides coroutines for structured concurrency, simplifying asynchronous programming by abstracting thread management. Crucially, Kotlin remains fully interoperable with Java, allowing classes from both languages to coexist and interact within the same project. While these features enhance productivity, they also introduce new language semantics and security mechanisms at the system level. According to Google, Kotlin is now ``\textit{used by over 60\% of professional Android developers}''~\cite{url_kotlin}.

Beyond third-party applications, Kotlin has seen increasing adoption within the Android OS itself. The Android Open Source Project (AOSP) team has been gradually migrating system apps and components from Java to Kotlin~\cite{url_aosp_to_kotlin}. Google’s official contribution guidelines now explicitly state that parts of the AOSP codebase are written in Kotlin and accept submissions in this language~\cite{url_Submit_code_changes}.

\begin{figure}[t]
	\centering
	\includegraphics[width=1\columnwidth]{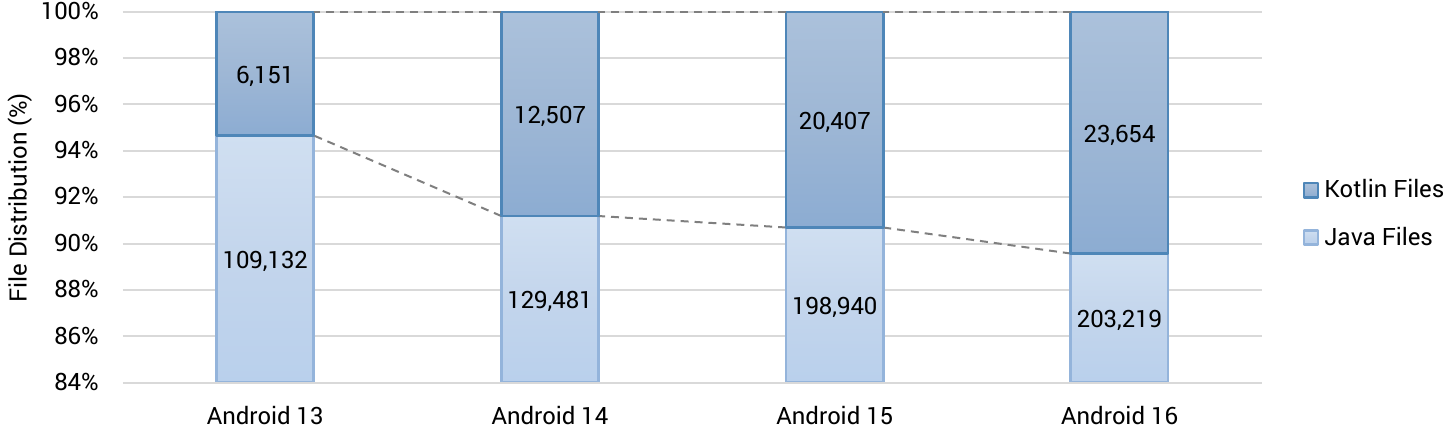}
	\caption{Trend of Kotlin and Java file counts in AOSP.}
	\label{fig:java-kotlin}
\end{figure}

To quantify the accelerating adoption of Kotlin within the OS framework, we conducted a longitudinal tracking of codebase compositions across four recent major releases: Android 13 through 16 (specifically build tags \texttt{13.0.0\_r70}, \texttt{14.0.0\_r37}, \texttt{15.0.0\_r34}, and \texttt{16.0.0\_r3}). As illustrated in Figure~\ref{fig:java-kotlin}, while Java remains the dominant language, the volume of Kotlin files has surged from 6,151 to 23,654, with the proportion nearly doubling from 5.3\% to 10.4\%. This trajectory underscores the deepening integration of Kotlin into core AOSP components.

\subsection{Threat Model}
\label{subsec:threat_model}

We consider an adversary who can execute code on a victim Android device through some local or remote means, for example, by installing a malicious app via official markets or third-party channels. The adversary aims to circumvent Android's security mechanisms, including permission enforcement, sandboxing, and other access control checks, to cause unauthorized behaviors such as privilege escalation or sensitive data leakage.

In this work, we specifically focus on vulnerabilities exposed by the \textit{clues derived from behavioral discrepancies} between the Kotlin and Java implementations of identical AOSP components. It is important to note that this threat model specifies the attacker’s capability and the security impact, rather than guiding the design of \toolname. While \toolname identifies potentially risky divergences, the threat model serves as the basis for determining their actual exploitability.

\section{Motivating Case and Challenges}
\label{sec:security}

This section presents a motivating case showing how discrepancies in parallel implementations can be exploited, followed by challenges in systematic analysis.

\subsection{Motivating Case}
\label{subsec:motivation-case}

To motivate our study, we examine a concrete instance from the Android framework where parallel implementations of the same functionality lead to security-relevant inconsistencies. In Android 14, we observed that certain system components are implemented in both Java and Kotlin. While there is no official documentation explaining this redundancy, our analysis indicates it is likely a byproduct of a staged refactoring from Java to Kotlin. This is supported by source code comments, such as: ``\texttt{\textit{/** Modern implementation of [PermissionManagerServiceInterface]. */}}''~\cite{url_PermissionService_kt}.

A representative example is the permission management service, which features two parallel implementations: the Java-based \texttt{PermissionManagerServiceImpl.java}~\cite{url_PermissionManagerServiceImpl_java} and the Kotlin-based \texttt{PermissionService.kt}~\cite{url_PermissionService_kt}. These classes expose overlapping entry points, with the active path selected at boot time based on the Android version (controlled by \texttt{SdkLevel.isAtLeastV()}, checking if the device is running on Android 15 or newer). Devices running Android 15 or newer utilize the Kotlin implementation, while Android 14 retains the Java version. Despite being conceptually equivalent, these two paths differ significantly in enforcement logic.

\paragraphNew{Semantic Divergence} We identified a semantic divergence in the \texttt{getAppOpPermissionPackages} method. This method returns packages that request a specific \textit{app-op} permission, where app-ops (App Operations) provide finer-grained access control than the traditional permission model.

In the Java implementation (Listing~\ref{code:getAppOp_Java}), the method acts defensively. It verifies whether the specified permission is present in the app-op registry. If the permission does not map to a valid \texttt{app-op}, the internal registry returns \texttt{null}, causing the method to strictly return an empty array. This safeguards the API from leaking data regarding unrelated permission types.

\begin{lstlisting}[language=Java, 
	style=mystyle,
    literate={@}{{@}}1,
	basicstyle=\ttfamily\footnotesize, 
	caption={Java version of \texttt{getAppOpPermissionPackages}.}, 
	label={code:getAppOp_Java}]
public String[] getAppOpPermissionPackages(@NonNull String permissionName) {
    Objects.requireNonNull(permissionName, "permissionName");
    return PermissionManagerServiceImpl.this.getAppOpPermissionPackagesInternal(permissionName);
}
    
private String[] getAppOpPermissionPackagesInternal(@NonNull String permName) {
  synchronized (mLock) {
    final ArraySet<String> packageNames = mRegistry.getAppOpPermissionPackages(permName);
        
    if (packageNames == null) {
      return EmptyArray.STRING; 
    }
    return packageNames.toArray(new String[0]);
  }
}
\end{lstlisting}

In contrast, the Kotlin implementation (Listing~\ref{code:getAppOp_Kotlin}) attempts to replicate this logic but introduces a control-flow flaw. The conditional check identifies non-app-op permissions but fails to halt execution (missing \texttt{return}, \textsf{Lines 7-9}), allowing the control flow to fall through. Consequently, it returns the package list requesting \textit{any} permission type, not just app-ops.

\begin{lstlisting}[language=Java, 
	style=mystyle,
    literate={@}{{@}}1,
    mathescape=false,
	basicstyle=\ttfamily\footnotesize, 
	caption={Kotlin version of \texttt{getAppOpPermissionPackages}.}, 
	label={code:getAppOp_Kotlin}]
override fun getAppOpPermissionPackages(permissionName: String): Array<String> {
  ...
  val permission = service.getState { 
    with(policy) { getPermissions()[permissionName] } 
  }
    
  if (permission == null || !permission.isAppOp) {
    packageNames.toTypedArray() 
  }
  ...
  return packageNames.toTypedArray()
}
\end{lstlisting}

\paragraphNew{Vulnerability and Impact} This behavioral difference causes unauthorized information disclosure. The Java version correctly hides usage data for non-app-op permissions (e.g., \texttt{RECEIVE\_SMS}). Conversely, the Kotlin version permits callers to query \textit{any} permission and enumerate all apps holding it.

We confirmed this control-flow flaw by invoking the Kotlin implementation in Android 15 and 16. Although this method is not directly accessible via the public Android SDK, it can be triggered via the Android Debug Bridge (\texttt{adb}) using a \texttt{service call}. We supplied a non-app-op system permission name, bypassed the intended visibility restriction, and enumerated apps successfully. We reported this issue to the Android Security Team. It was acknowledged and classified as \textbf{Low Severity} (requiring local shell access), nevertheless, it provides concrete evidence that semantic gaps with security risk exist between Java-Kotlin parallel implementations.

\paragraphNew{Summary} 
This motivating case highlights a critical observation: Java-Kotlin parallel implementations in AOSP Android are rarely straightforward translations. In practice, system developers often introduce subtle semantic deviations for reasons such as logic optimization, bug fixes, or the adoption of new language features. These differences can serve as valuable clues that may reveal underlying security flaws in the complex and extensive Android system.

However, identifying these discrepancies is non-trivial. A robust analysis must extend beyond individual methods to examine entire call chains and their interactions. Furthermore, the significant divergence in syntax and coding idioms between Java and Kotlin makes manual comparison prohibitively expensive and error-prone. Consequently, an automated approach is essential. Motivated by these findings, this work conducts a systematic study using \toolname to identify and analyze these parallel implementations at scale.

\subsection{Challenges for Automated Analysis}
\label{subsec:challenges}

To systematically assess the security implications of parallel implementations in AOSP, our goal is to (1) automatically identify all parallel pairs of Java and Kotlin implementations, and (2) detect potentially risky behavioral discrepancies between them. Achieving this requires addressing two key challenges.

\paragraphNew{Challenge 1: Identifying Parallel Implementations} 
Locating parallel implementations is non-trivial because they could reside in different packages, employ distinct naming conventions, or evolve independently. Furthermore, the syntactic gap between Java and Kotlin, ranging from variable declarations to inheritance models, prevents direct source-level comparison. To our knowledge, there is no practical tool that can convert between Java and Kotlin source code while strictly preserving semantic equivalence. Additionally, differences in language rules, such as method overriding, return type covariance, and nullability constraints, further complicate the mapping process. Consequently, identifying parallel implementations within the massive AOSP codebase poses a significant scalability and accuracy challenge.

\paragraphNew{Solution to Challenge 1} 
We address this challenge through a three-step process involving bytecode normalization, context recovery, and signature-based identification. We explicitly avoid fuzzy name matching because AOSP contains widespread method name duplication across unrelated classes. Relying on loose matching would introduce excessive false positives and incur prohibitive manual and computational costs.

\miniparagraph
\textit{Step 1: Normalizing method declarations.} Both Java and Kotlin in AOSP compile into Dalvik bytecode. Extracting DEX files from the compiled AOSP yields a uniform bytecode-level IR. This abstraction neutralizes language-specific syntactic divergences and allows for a unified analysis of the entire framework.

\miniparagraph
\textit{Step 2: Reconstructing class-to-source mappings.} Compilation obscures the direct link between classes and source files. However, \texttt{userdebug} and \texttt{eng} builds retain crucial debugging metadata. Parsing this metadata from DEX files accurately reconstructs mappings between compiled classes and their original source files. This reconstruction enables tracing bytecode back to its definition.

\miniparagraph
\textit{Step 3: Identification via migration patterns.} After establishing class-source associations, we compare method signatures to locate parallel candidates. A signature comprises the method name, parameter types, and return type. Because new implementations must preserve original API contracts for backward compatibility, classes with substantial signature overlap qualify as candidates. Manually inspecting file paths, commit logs, and documentation reveals two primary parallel implementation types (Statically Observable and Runtime-Gated), as detailed in Section~\ref{subsec:Methods-Identification}. \toolname uses these patterns to identify parallel pairs with high precision.

\paragraphNew{Challenge 2: Detecting Security-Critical Discrepancies}
Parallel implementations aim for functional equivalence, so their high-level semantics are generally aligned. However, subtle code-level deviations can introduce inconsistent behaviors that pose security risks. Detecting such risky discrepancies is difficult because it requires determining: (1) \textit{where} the execution logic diverges, and (2) \textit{whether} that divergence compromises the security posture.

Traditional program analysis techniques, such as AST or CFG comparison, struggle in this context because Java and Kotlin implementations often diverge significantly in structure and coding idioms. For example, Kotlin's use of extension functions or null-safety sugar creates syntactic disparities. These disparities create a wide structural gap that renders standard isomorphism checks ineffective, often leading to excessive false positives where code looks different but acts the same.

\paragraphNew{Solution to Challenge 2}
To address this challenge, \toolname replaces direct raw-code and rigid AST comparisons with a structured hybrid approach. First, we introduce a language-agnostic shallow parser to standardize heterogeneous Java and Kotlin methods into a \textit{Unified Execution Graph (UEG)}. This representation preserves core execution and data-flow logic while removing language-specific syntactic noise. 

Building upon the UEG, we leverage LLMs to perform cross-graph differential analysis. The LLM performs \textit{semantic N:M mapping} to dynamically align execution paths. It explicitly pinpoints logical deviations, such as bypassed guards or fail-open fall-throughs, and extracts them into a \textit{Structured Divergence Record (SDR)}. Finally, \toolname employs a Contextual Verification and Retrieval-Augmented Generation (RAG) mechanism grounded in official Android severity criteria. An ensemble majority-voting strategy complements this process for assessing security risks and suppressing hallucinations. This defense-in-depth pipeline ensures the system reports only genuine, exploitable vulnerabilities.
\section{Design of \toolname}
\label{sec:design}

\begin{figure*}[t]
	\centering
	\includegraphics[width=1\linewidth]{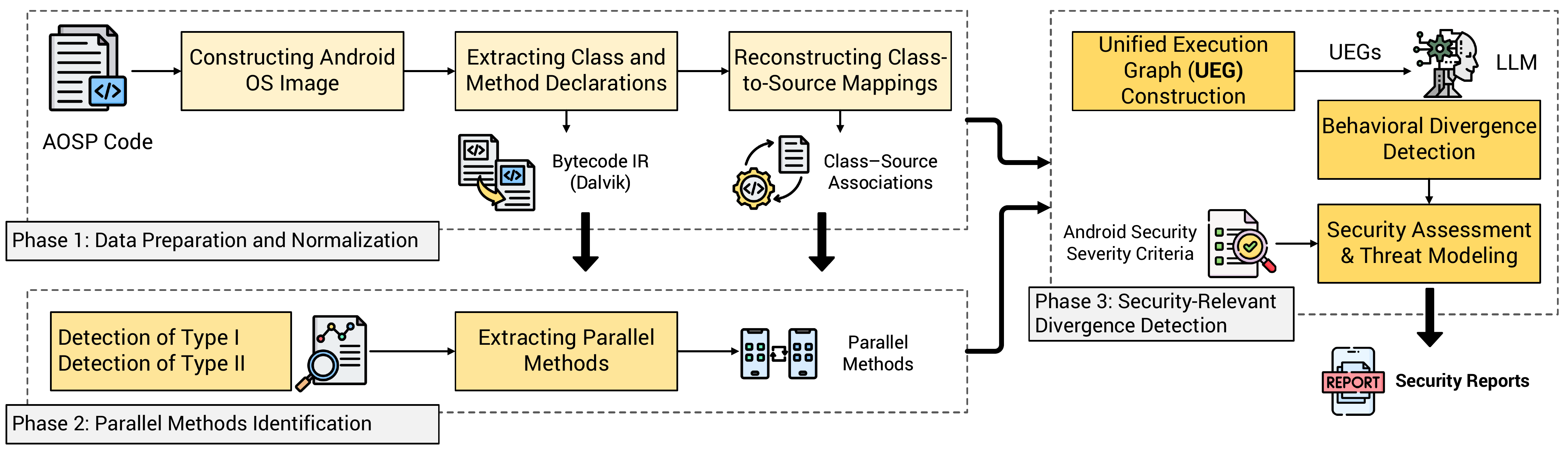}
	\caption{Overview of \toolname.}
	\label{fig:overview}
\end{figure*}

Based on the solutions to challenges outlined in Section~\ref{sec:security}, we design and implement \toolname, an automated framework for systematically analyzing parallel Java-Kotlin implementations in the Android system. \toolname operates by identifying parallel components at scale and performing differential analysis to uncover security-relevant discrepancies. As illustrated in Figure~\ref{fig:overview}, the workflow consists of three distinct phases:

\begin{description}[leftmargin = 15pt, topsep=2pt]
    \item[\textbf{\textit{Phase 1:}}] \textit{\textbf{Data Preparation and Normalization}}. This phase compiles the Android codebase and normalizes heterogeneous source code into a unified intermediate representation to facilitate cross-language analysis.

    \item[\textbf{\textit{Phase 2:}}] \textit{\textbf{Parallel Method Identification}}. This phase identifies candidate class pairs and extracts parallel method information across the Java-Kotlin boundary.

   \item[\textbf{\textit{Phase 3:}}] \textit{\textbf{Security-Relevant Divergence Detection}}. This phase converts relevant Java and Kotlin code into a unified structured representation. The system then leverages LLM-based reasoning to detect semantic divergences between parallel methods and assess their potential to introduce security vulnerabilities.
\end{description}

\subsection{Data Preparation and Normalization}
\label{subsec:preprocessing}

To ensure accurate downstream analysis, \toolname first transforms the Android source tree into a normalized state that supports reliable class-to-source mapping. This step is foundational, as it bridges the gap between compiled artifacts and source code, enabling a precise distinction between Java and Kotlin components.

\paragraphNew{Constructing Android OS Image}
The analysis begins with the Android Open Source Project (AOSP). To obtain the ground truth of the system's execution logic, \toolname compiles the full AOSP source tree. This produces a system image containing the authentic compiled classes (DEX files) used by the Android framework, ensuring our analysis reflects the actual runtime behavior.

\paragraphNew{Extracting Class and Method Declarations}
To enable cross-language analysis, \toolname abstracts away language-specific syntax by lifting both Java and Kotlin into a common format. It extracts DEX files from the compiled system image and parses the bytecode to reconstruct class and method declarations. These declarations are then normalized into a unified bytecode-level Intermediate Representation (IR). Note that while this low-level IR is crucial for precise signature matching in Phase 2, it lacks the high-level semantic context (e.g., original variable names, inline comments) necessary for LLM reasoning in Phase 3.

\paragraphNew{Reconstructing Class-to-Source Mappings}
Compilation typically severs the direct link between binary classes and their source files, which hinders the identification of parallel implementations. To resolve this, \toolname leverages the debugging metadata preserved in \texttt{userdebug} or \texttt{eng} builds. By parsing the class descriptor and source file attributes within the DEX files, \toolname accurately reconstructs the mapping between every compiled class and its original source definition. It then traverses the AOSP source tree to verify these associations, building a comprehensive index that links bytecode to specific Java or Kotlin files.

\subsection{Parallel Methods Identification}
\label{subsec:Methods-Identification}

Following normalization, \toolname proceeds to identify \textit{parallel methods} -- pairs of Java and Kotlin methods that implement identical functionality within the AOSP framework. We categorize these implementations into two distinct types based on their structural and runtime characteristics.

\paragraphNew{\textsf{Type I}: Statically Observable Migrations}
This category encompasses implementations that can be identified through file naming conventions and directory structures. These patterns typically reflect a staged migration strategy where the relationship between the legacy Java code and the new Kotlin code is explicitly encoded in the file system. We observe two common manifestations:

\miniparagraph
(1) \textit{Co-located Naming Variations:} The Java and Kotlin files reside in the same directory but are distinguished by a modified filename convention. AOSP frequently appends a numeric suffix to the Kotlin file (e.g., \texttt{WifiDialog2.kt}), but this pattern also covers other systematic renaming strategies used to avoid class name conflicts. This typically indicates a staged replacement. As shown in Listing~\ref{code:pattern1}, \texttt{WifiDialog.java}~\cite{url_WifiDialog_java} and \texttt{WifiDialog2.kt}~\cite{url_WifiDialog2_kt} coexist, and the Java documentation explicitly advises migration: ``\textit{this object will be removed in the near future, please develop in {\texttt{@link WifiDialog2}}}''. This confirms that they are parallel implementations of the Wi-Fi dialog UI.

\begin{lstlisting}[language=Java, 
	style=mystyle,
    literate={@}{{@}}1,
	basicstyle=\ttfamily\footnotesize, 
	caption={Example of \textsf{Type I} (Co-located Suffix).}, 
	label={code:pattern1}]
------------------------Java--------------------------
/packages/apps/Settings/src/com/android/settings/wifi/WifiDialog.java
/**
* Dialog for users to edit a Wi-Fi network
*
* Migrating from Wi-Fi SettingsLib to WifiTrackerLib, this object will be removed in the near future, please develop in {@link WifiDialog2}.
*/
public class WifiDialog {...}

-----------------------Kotlin-------------------------
/packages/apps/Settings/src/com/android/settings/wifi/WifiDialog2.kt
/**
* Dialog for users to edit a Wi-Fi network
*/
@OpenForTesting
open class WifiDialog2 {...}
\end{lstlisting}

\miniparagraph
(2) \textit{Directory Shadowing:} Java and Kotlin files share an identical filename but reside in parallel directory structures, distinguished by a Kotlin-specific segment in the Kotlin path, such as \texttt{/kotlin/}. This effectively creates a ``shadow'' implementation. As illustrated in Listing~\ref{code:pattern2}, both \texttt{AudioPreview.java}~\cite{url_AudioPreview_java} and \texttt{AudioPreview.kt}~\cite{url_AudioPreview_kt} provide the audio preview dialog. AOSP commit logs confirm this parallel existence and explicitly note the introduction of the Kotlin version alongside the original Java code (e.g.,``\textit{Add Kotlin version of Music Java code}''~\cite{url_Commit_c95a176}).

\begin{lstlisting}[language=Java, 
	style=mystyle,
    literate={@}{{@}}1,
	basicstyle=\ttfamily\footnotesize, 
	caption={Example of \textsf{Type I} (Directory Shadowing).}, 
	label={code:pattern2}]
------------------------Java--------------------------
/packages/apps/Music/src/com/android/music/AudioPreview.java
/**
* Dialog that comes up in response to various music-related VIEW intents.
*/
public class AudioPreview {...}

-----------------------Kotlin-------------------------
/packages/apps/Music/kotlin/src/com/android/music/AudioPreview.kt
/**
* Dialog that comes up in response to various music-related VIEW intents.
*/
class AudioPreview {...}
\end{lstlisting}

\paragraphNew{\textsf{Type II}: Runtime-Gated Migrations}
The second category is more subtle and cannot be detected via file metadata alone. Implementations may reside in disparate packages with no obvious naming correlation. The system dynamically selects the active implementation at runtime based on the OS version or feature flags.

Listing~\ref{code:pattern3} illustrates this within the permission subsystem mentioned in Section~\ref{subsec:motivation-case}, which co-hosts both the Java-based \texttt{Permissi- onManagerServiceImpl} and the Kotlin-based \texttt{PermissionService}. A version check (e.g., \texttt{SdkLevel.isAtLeastV()}) governs the control flow: newer Android versions retrieve the Kotlin implementation via the system service registry, while older versions fall back to instantiating the Java class directly.

\begin{lstlisting}[language=Java, 
	style=mystyle,
    literate={@}{{@}}1,
	basicstyle=\ttfamily\footnotesize, 
	caption={Example of \textsf{Type II}  (Runtime Switch).}, 
	label={code:pattern3}]
/frameworks/base/core/java/android/permission/PermissionManager.java
/**
* Whether to use the new {@link com.android.server.permission.access.AccessCheckingService}.
*/
public static final boolean USE_ACCESS_CHECKING_SERVICE = SdkLevel.isAtLeastV();
\end{lstlisting}

\paragraphNew{Detection of \textsf{Type I} (Static Patterns)}
\toolname identifies \textit{Statically Observable Migrations} using a two-step heuristic approach:

\miniparagraph
(1) \textit{Path and Filename Matching:} \toolname traverses the AOSP source tree to index all Java and Kotlin files. It generates candidate pairs based on naming conventions: (a) co-located files sharing a base name but distinguished by systematic variation (e.g., a numeric suffix), or (b) files with identical names where one resides within a Kotlin-specific subdirectory.

\miniparagraph
(2) \textit{Method Signature Overlap Check:} \toolname validates each candidate pair using the class-to-source mappings established in Phase~1. It retrieves the compiled definitions for the corresponding Java and Kotlin classes and compares their method signatures. \toolname classifies the pair as a parallel implementation if they share at least one method signature. Although the naming conventions in \textsf{Type I} are strong indicators, this content-based check is necessary. It serves as a safeguard to filter out false positives in which files share a name but perform unrelated tasks.

\vspace{2pt}
\noindent\textit{Handling Build Conflicts for \textsf{Type I} (Directory Shadowing):} A specific challenge arises with \textsf{Type I} (Directory Shadowing) because the Java and Kotlin files define identical class names. Consequently, they cannot coexist in a single build. Standard AOSP build configurations typically include only one version at a time. To analyze both, \toolname performs a separate compilation step. We adjust the build configuration to select the other implementation as the active one. This ensures that the previously excluded file is compiled and available for analysis. We execute this alternate compilation once after the initial file scanning.

\paragraphNew{Detection of \textsf{Type II} (Runtime Patterns)}
Detecting \textit{Runtime-Gated Migrations} requires analyzing control flow logic. \toolname employs targeted data-flow analysis:

\miniparagraph
(1) \textit{Version-Gated Branch Tracing:} \toolname first scans the AOSP codebase to identify OS version check operations. These include API calls matching the pattern \texttt{SdkLevel.isAtLeast*} and comparisons involving \texttt{Build.VERSION.SDK\_INT}. For each identified check, the tool performs data-flow analysis to track how the boolean result propagates through assignments, method calls, fields, and return values. When this result determines the path of an \texttt{if/else} statement, \toolname inspects both branches. It records the specific class that is instantiated or retrieved within each branch.

\miniparagraph
(2) \textit{Matching Java/Kotlin Pairs:} \toolname uses the class-to-source associations to determine the implementation language of the classes found in the previous step. It identifies instances where one branch loads a Kotlin class and the other loads a Java class. If these two classes share overlapping method signatures, the tool marks them as a parallel pair under \textsf{Type II}.

\vspace{3pt}
By applying these techniques, \toolname identifies the final set of parallel class pairs.

\paragraphNew{Extracting Parallel Methods}
For each identified Java-Kotlin class pair, \toolname extracts the methods that share identical signatures. We refer to these as \textit{parallel methods}. They form the basis for our subsequent behavioral comparison.

\subsection{Security-Relevant Divergence Detection}
\label{subsec:detection}

The final phase analyzes the parallel methods to identify behavioral discrepancies that pose security risks. It subsequently generates a comprehensive report. To bridge the semantic gap between Java and Kotlin while maintaining analytical rigor, \toolname avoids direct comparisons of raw source code. Instead, the system standardizes the heterogeneous inputs into a unified structured representation and employs a multi-stage, structured LLM reasoning pipeline.

\paragraphNew{Code Context Extraction and UEG Construction}
\toolname first isolates the code context required for analysis. For each parallel method, it parses the complete call chain and all invoked sub-routines. Using the class-to-source mappings from Phase 1, \toolname retrieves the source code for the entry method and its dependencies for downstream processing.

The Phase 1 bytecode IR effectively matches signatures but discards critical developer intent (e.g., inline comments, variable naming). It remains too verbose for LLMs to analyze high-level logic. Conversely, supplying raw source text introduces excessive cross-language syntactic noise. This noise inevitably undermines security analysis precision. Furthermore, traditional static analyzers (e.g., Soot or WALA) relying on Abstract Syntax Trees (AST) frequently fail or crash on isolated, partially unresolvable framework snippets.

To resolve these issues, we introduce a language-agnostic shallow parser that bypasses the full AST generation step. It processes the extracted Java and Kotlin source streams uniformly to transform heterogeneous code pairs into a \textit{\textbf{Unified Execution Graph (UEG)}}. This standardized representation preserves core execution logic through a three-phase pipeline:

\miniparagraph
(1) \textit{Lexical Preprocessing \& Normalization:} To avoid language-specific dependencies, \toolname applies a token-level filter to strip non-executable elements. This includes comments, annotations, and string literals. This sanitization ensures that downstream topological extraction operates exclusively on executable logic and remains immune to textual noise.

\miniparagraph
(2) \textit{Intra-Procedural CFG Construction:} \toolname aggregates sequential statements into logical basic blocks (BBs) via scope depth tracking and control-flow keyword recognition. It then computes directed control-flow graphs (\texttt{cfg}) by tracing branch conditions and resolving post-dominator nodes for path convergence. The pipeline performs control-flow normalization by mapping language-specific constructs, such as Kotlin's non-local returns (\texttt{return@label}) and Elvis operators (\texttt{?:}), into standard execution bounds equivalent to Java control flows.

\miniparagraph
(3) \textit{Conservative Call-Edge Resolution:} To capture inter-procedural behaviors, \toolname extracts explicit function invocations (\texttt{calls}) within each basic block. Since the shallow parser lacks the deep type inference needed to perfectly resolve polymorphic dispatch across different languages, it employs a weakly typed over-approximation strategy. Call sites are resolved to a candidate pool of all matching overloaded signatures. This approach preserves execution completeness and prevents analysis crashes.

The resulting UEG is a deterministic, language-neutral JSON topology consisting of method signatures, basic block arrays (\texttt{cfg}), and intra-block invocation edges (\texttt{calls}). This format ensures the downstream LLM analyzes a pure graph representing execution and data flow. This representation is entirely free of Java and Kotlin syntactic disparities. Listing~\ref{code:ueg} provides an example.

\begin{lstlisting}[language=json, 
 style=mystyle,
 literate={@}{{@}}1,
 basicstyle=\ttfamily\footnotesize,
 breaklines=true,
 caption={Partial UEG for \texttt{getAppOpPermissionPackages} (Kotlin version).}, 
 label={code:ueg}]
{
 "kotlin_implementation": {
  "PermissionService.getAppOpPermissionPackages (Entry)": {
   "sig": "<fun getAppOpPermissionPackages(String) : Array<String>>",
   "calls": {
    "BB_2": [ "AccessCheckingService.getState (Callee)" ],
    "BB_4": [ "AppIdPermissionPolicy.getPermissions (Callee)" ]
   },
   "cfg": {
    "BB_0": ["override fun getAppOpPermissionPackages(permissionName: String): Array<String> {\nrequireNotNull(permissionName) {", [ "BB_1" ]],
    "BB_1": ["\"permissionName cannot be null\"", [ "BB_2" ]],
    "BB_2": ["val packageNames = ArraySet<String>()\nval permission = service.getState {", [ "BB_3" ]],
    "BB_3": ["with(policy) {", [ "BB_4" ]],
    "BB_4": ["getPermissions()[permissionName]", [ "BB_5" ]],
    "BB_5": ["if (permission == null || !permission.isAppOp) {", [ "BB_6", "BB_7" ]],
    "BB_6": ["packageNames.toTypedArray()", [ "BB_7" ]],
    "BB_7": ["packageManagerLocal.withUnfilteredSnapshot().use { snapshot ->", [ "BB_8" ]],
    "BB_8": ["snapshot.packageStates.forEach packageStates@{ (_, packageState) ->", [ "BB_9", "BB_12" ]],
    "BB_9": ["val androidPackage = packageState.androidPackage ?: return@packageStates", [ "BB_10", "BB_8" ]],
    "BB_10": ["if (permissionName in androidPackage.requestedPermissions) {", [ "BB_11", "BB_8" ]],
    "BB_11": ["packageNames += androidPackage.packageName", [ "BB_8" ]],
    "BB_12": ["return packageNames.toTypedArray()", [ "End" ]]
   }
  },
  ... // Omit other parts
 }
}
\end{lstlisting}

\paragraphNew{Behavioral Divergence Detection}
Following UEG normalization, \toolname performs a comparative analysis. We prompt the LLM, acting as a Principal Android OS Engineer, to identify structural and behavioral divergences.

During this phase, the LLM traverses the paired UEGs, tracking cross-boundary data propagation via \texttt{cfg} and \texttt{calls} edges to align execution paths. To handle migration-induced structural shifts, \textit{semantic N:M mapping} guides the LLM to group related blocks by computational intent rather than syntactic boundaries. Since refactoring often alters method boundaries, the LLM performs a cross-boundary search across the entire \texttt{cfg} pool rather than restricting matches to identically named methods.

Guided by this alignment, the LLM extracts semantic signals across four computational deviations. First, \textit{guard \& state integrity} identifies discrepancies where one implementation enforces strict admission checks before state mutations while the other bypasses them. Second, \textit{fail-open fall-throughs} detect explicit aborts like \texttt{throw} or \texttt{return} reduced to non-blocking operations. Third, \textit{data resolution \& terminal state asymmetry} distinguishes outcomes derived from static scalars versus dynamic runtime evaluations. Fourth, \textit{core logic deviations} capture misaligned mathematical operations, differing loop boundaries, or altered exception-handling ranges.

Finally, \toolname constrains the LLM to output a \textit{\textbf{Structured Divergence Record (SDR)}} in JSON format for each deviation, as the example in Listing~\ref{code:sdr}. This object contains a reasoning trace documenting call chains, mapped blocks, and extracted guard conditions. The LLM grounds every discrepancy description purely in observable code facts. Each explanation is tightly anchored to specific basic block IDs, such as \texttt{BB\_4} in Java versus \texttt{BB\_6} in Kotlin. These explicit coordinates provide traceable evidence for the subsequent validation phase.

\begin{lstlisting}[language=json, 
 style=mystyle,
 literate={@}{{@}}1,
 basicstyle=\ttfamily\footnotesize,
 breaklines=true,
 caption={Partial SDR for \texttt{getAppOpPermissionPackages}.}, 
 label={code:sdr}]
{
 "divergence_id": "div_2",
 "analyzed_java_entry_method": "public String[] getAppOpPermissionPackages(@NonNull String permissionName)",
 "analyzed_kotlin_entry_method": "override fun getAppOpPermissionPackages(permissionName: String): Array<String>",
 "java_path_involved": [
  "PermissionManagerServiceImpl.getAppOpPermissionPackagesInternal::BB_1",
  "PermissionManagerServiceImpl.getAppOpPermissionPackagesInternal::BB_2"
 ],
 "kotlin_path_involved": [
  "PermissionService.getAppOpPermissionPackages::BB_5",
  "PermissionService.getAppOpPermissionPackages::BB_6",
  "PermissionService.getAppOpPermissionPackages::BB_7"
 ],
 "neutral_divergence_description": "Fail-Open Guard Logic: In Java, if the permission is not present in the AppOp registry, it returns 'EmptyArray.STRING' immediately (BB_2). In Kotlin, the check 'if (permission == null || !permission.isAppOp)' (BB_5) is effectively ignored; the flow proceeds from BB_6 into the full package scan (BB_7), returning packages that request the name regardless of its AppOp status."
}
\end{lstlisting}

\paragraphNew{Security Assessment \& Threat Modeling}
Not all behavioral divergences translate to security vulnerabilities. To filter out benign variations and objectively evaluate risks, \toolname initiates a two-step Security Assessment phase for the generated SDRs.

\miniparagraph
(1) \textit{Contextual Consistency Verification:} The LLM first validates SDR authenticity to prevent hallucinations by querying the provided block IDs against the original UEG. This confirms whether the semantic intent genuinely changed. The LLM discards the divergence if architectural refactoring merely moved the logic to a deeper sub-callee or if structural shifts do not affect the actual data flow.

\miniparagraph
(2) \textit{RAG-Enhanced Threat Modeling \& Assessment:} For verified divergences, the model evaluates if the discrepancy constitutes a newly introduced Kotlin regression or an unpatched legacy Java flaw. To assess exploitability and prevent subjective overestimation, \toolname utilizes a Retrieval-Augmented Generation (RAG) mechanism. This mechanism injects the official Android Security Severity Criteria~\cite{url_Severity} into the LLM context. The augmented LLM systematically analyzes divergence across core threat-modeling dimensions. These dimensions span unprivileged IPC reachability, the integrity of global OS states, internal metadata exposure, and critical execution anomalies like TOCTOU races or system-wide DoS. By grounding its reasoning in these unified rules and official metrics, the LLM maps the vulnerability to an objective severity tier (Critical, High, Moderate, Low, or None) and discards issues with negligible impacts. 

Following the assessment, \toolname outputs a security assessment report detailing the severity reasoning and exploit rationale, as shown in Listing~\ref{code:report}. To mitigate LLM non-determinism, the system determines the final evaluation via majority voting (2/3 consensus).

\begin{lstlisting}[language=json, 
 style=mystyle,
 literate={@}{{@}}1,
 basicstyle=\ttfamily\footnotesize,
 breaklines=true,
 caption={Partial report for \texttt{getAppOpPermissionPackages}.}, 
 label={code:report}]
{
 "is_vulnerable": true,
 "affected_implementation": "Kotlin (Regression)",
 "vulnerability_type": "InfoLeak",
 "severity_rating": "Low",
 "severity_reasoning": "Leaking sensitive system metadata, orphaned states, or historical app footprints.",
 "exploit_rationale": "An attacker queries 'getAppOpPermissionPackages' with a known non-AppOp permission string (e.g., a custom permission exclusively used by a targeted banking or social app). Because the Kotlin implementation lacks the 'return' keyword during its validation phase, the early-exit check fails open. The OS then actively scans the unfiltered snapshot and returns the names of all packages declaring that permission. This allows an unprivileged attacker to reliably enumerate installed packages on the device, fully bypassing Package Visibility (QUERY_ALL_PACKAGES) sandbox restrictions."
}
\end{lstlisting}

\paragraphNew{Remarks: Systematic Hallucination Mitigation}
Applying LLMs to complex program analysis risks producing speculative or fabricated findings. To ensure highly reliable vulnerability reports, \toolname employs a five-layer defense-in-depth mitigation strategy. (1) \textit{Structural Grounding} restricts the LLM to the language-agnostic UEG rather than raw source code. This removes syntactic ambiguities that often mislead generative models. (2) \textit{Constrained Chain-of-Thought (CoT)} requires the output of intermediate reasoning states, such as the \texttt{reasoning\_trace} in the SDR. The LLM must explicitly document call stacks, block mappings, and unrolled guard conditions before formulating descriptions. This anchors deductions in observable graph facts instead of latent textual bias. (3) \textit{Anchor-Based Contextual Verification} limits the LLM to a targeted inspection bounded by basic block IDs in the SDR. Verifying these anchors against entry methods effectively constrains the generation space. (4) \textit{RAG-Grounded Threat Modeling} injects the official Android Security Severity Criteria as an external source of truth. This prevents the model from subjectively inventing threat scenarios by forcing it to evaluate exploitability against objective rules. (5) \textit{Ensemble Consensus} executes three independent parallel inferences during threat modeling. A strict majority-voting mechanism requires at least a 2/3 consensus on both the \texttt{"is\_vulnerable"} boolean and the severity rating. This statistical agreement neutralizes non-deterministic outliers and single-run errors, ensuring the system reports only reproducible vulnerabilities.

\section{Evaluation and Results}
\label{sec:results}

We evaluated \toolname on AOSP to assess the security implications of parallel implementations. Specifically, this evaluation focuses on the following three research questions.

\begin{mybox}[boxsep=0pt,
	boxrule=1pt,
	left=4pt,
	right=4pt,
	top=4pt,
	bottom=4pt,
	colback=backcolor]
	
\begin{description}[leftmargin=12pt, topsep=-1pt, itemsep=-1pt]

\item [\textbf{RQ1:}] \textit{How sound is the workflow design of \toolname?}

\item [\textbf{RQ2:}] \textit{What empirical findings does \toolname reveal within the AOSP codebase?}

\item [\textbf{RQ3:}] \textit{What are the practical security implications of the discovered vulnerabilities?}

\end{description}
\end{mybox}

\subsection{Implementation and Experiment Setup} 

\paragraphNew{Prototype Implementation}
We implemented a prototype of \toolname comprising 7,092 lines of code (2,761 in Java, 2,674 in Python, and 1,657 in JavaScript). \toolname integrates \texttt{dexdump} for DEX parsing and \texttt{Soot} for bytecode analysis, while leveraging \texttt{javalang} and \texttt{kopyt} for source parsing.

\paragraphNew{Experimental Data}
We conducted our evaluation using the AOSP codebases for Android 14, 15, and 16. The specific build tags are \texttt{14.0.0\_r37} (141,988 Java \& Kotlin files), \texttt{15.0.0\_r34} (219,347 files), and \texttt{16.0.0\_r3} (226,873 files). We compiled \texttt{userdebug} images for each version. This compilation target preserves the debugging metadata necessary for accurate class-to-source mapping.

Additionally, we manually constructed a \textit{ground truth dataset} for LLM selection and ablation experiments. This dataset contains 20 selected parallel method pairs. It encompasses 51 verified behavioral divergences. 9 of these cases are confirmed as security-relevant.

\paragraphNew{Execution Environment}
We conducted the experiment on an Ubuntu~22.04 server equipped with a 96-core AMD EPYC 9654 CPU and 1\,TB of RAM.

\paragraphNew{Usage of LLM}
This work utilized the \textsf{Gemini 3.1 Pro} model via the official Python SDK and implemented parallelization to improve query throughput. Preliminary tests against \textsf{GPT-5.4} and \textsf{Claude Opus 4.7} motivated this model selection. Evaluations on the ground-truth dataset demonstrated that \textsf{Gemini 3.1 Pro} achieved the best performance. Appendix~\ref{appendix:llm-select} details the specific experimental configurations and results. The model likely includes Android codebases in its training data. This exposure improves its understanding of framework code.

To reduce costs and comply with rate limits~\cite{url_Rate_limits}, we minimized the input context by retaining only the logic relevant to behavioral equivalence:

\begin{itemize}[leftmargin = 15pt, topsep=2pt]
\item We omit boilerplate and utility functions whose internal details are unnecessary for judging equivalence. Examples include \texttt{add}, \texttt{remove}, \texttt{toString}, and \texttt{println}.

\item If both implementations invoke the same callee, we exclude it from the context. A shared callee inherently does not introduce behavioral divergence.
\end{itemize}

\paragraphNew{Performance} 
On average, the end-to-end analysis for a single Android firmware image completed in 5.2 hours. The preliminary stages were executed efficiently. Phase 1 (Data Preparation and Normalization) and Phase 2 (Parallel Method Identification) required 57 minutes and 25 minutes, respectively. Phase 3 (Security-Relevant Divergence Detection) accounted for most of the runtime. Within this phase, the LLM-based steps consumed around 3.8 hours. The Code Context Extraction and UEG Construction required less than 10 seconds, introducing negligible overhead. The extended duration of Phase 3 stems primarily from LLM API round-trip latency and batching overhead. The average LLM usage per image was 5.9 million tokens.

\subsection{Ablation Experiments}
\label{subsec:ablation}

To evaluate the contribution of key Phase 3 components to the overall effectiveness of \toolname, we conducted ablation experiments.  We specifically analyze UEG Construction, Contextual Consistency Verification, and RAG Enhancement.

\paragraphNew{Impact of UEG on Behavioral Divergence Discovery} 
The first experiment evaluates whether the UEG representation improves behavioral divergence identification compared to raw source code. We compared the full \toolname pipeline against a source-code baseline. In real-world code, the population of non-divergent points is practically infinite and impossible to label exhaustively. Consequently, true negatives remain undefined. This renders recall and F1-score inapplicable for the discovery phase. We instead measure \textit{Coverage} (the percentage of identified ground-truth divergences) and \textit{Accuracy} (the percentage of genuine reported divergences). 

The UEG-based approach identified 47 behavioral divergences, achieving 80.4\% coverage and 87.2\% accuracy. The baseline identified 58 cases with 64.7\% coverage and 56.9\% accuracy. These findings confirm that the UEG effectively suppresses syntactic noise and standardizes cross-language control flows. It enables the LLM to focus on core semantic logic.

\paragraphNew{Impact of Verification and RAG on Security Assessment}
The second experiment isolates the impact of Contextual Consistency Verification and RAG Enhancement during security assessment. We used the 47 divergences identified in the previous experiment as a fixed set of inputs. This experiment constitutes a binary classification task over a closed set. We therefore report standard precision, recall, and F1-score.

The result (Table 1) shows that disabling these components
caused significant performance degradation. Removing the verification step led to a surge in false positives. The LLM could no longer cross-check findings against the deterministic UEG. Removing the RAG module caused the LLM to output subjective and inconsistent risk ratings.

\begin{table}[t]
\centering
\caption{Ablation experiment for security assessment.}
\label{tab:ablation}
\small \vspace{-5pt}
\begin{tabu} to \linewidth {*X[0.8,l]* X[0.4,c] X[0.5,c] | X[0.4,c]  X[0.5,c] | X[0.6,c] X[0.7,c]*}
\thickhline
\textbf{Steps}  &   \textsf{F1}  & \textsf{F1$\downarrow$}  & \textsf{Recall}  & \textsf{Recall$\downarrow$} &  \textsf{Precision}  & \textsf{Precision$\downarrow$}\\
\thickhline
Baseline & 15.38\% & 54.62\% & 25.00\% &  33.33\% & 11.11\% &  76.39\%\\ 
\hline
(w/o) Verif  & 17.39\% & 52.61\% & 40.00\% &  18.33\% & 11.11\% &  76.39\%\\ 
\hline
(w/o) RAG & 28.57\% & 41.43\% & 23.53\% &  34.80\% & 36.36\% &  51.14\%\\ 
\thickhline
\toolname & 70.00\% & - & 58.33\% &  - & 87.50\% &  -\\
\thickhline
\end{tabu}
\end{table}

\subsection{Experimental Results on AOSP} 
\label{subsec:results}
Based on the above experiment setup, we applied \toolname to analyze the AOSP Android 14-16 codebases for parallel implementations and their security implications.

\begin{table}[t]
\centering
\caption{Summary of analysis results by Android OS.}
\label{tab:results}
\small \vspace{-5pt}
\begin{tabu} to \linewidth {*X[1.2,l]* X[0.5,c] X[0.5,c] | X[0.9,c] | X[0.9,c]*}
\thickhline
\multirow{2}{*}{\textbf{Android Images}} & \multicolumn{2}{c|}{\textbf{Method Pairs}} &  \textbf{Behavioral} &    \textbf{Vulnerable} \\
  &   \textsf{Type I}  & \textsf{Type II}  & \vspace{-7pt}\textbf{Divergences}& \vspace{-7pt}\textbf{Divergences} \\
\thickhline
\texttt{14.0.0\_r37} & 81 & 70 & 189 &  20 \\ 
\hline
\texttt{15.0.0\_r34} & 54 & 71 & 158 &  17 \\ 
\hline
\texttt{16.0.0\_r3} & 53 & 0 & 25 &  0 \\ 
\thickhline
\textbf{Total} & \multicolumn{2}{c|}{\textbf{329}} & \textbf{372} &  \textbf{37} \\

\thickhline
\end{tabu}

\end{table}

\paragraphNew{Results Summary} As summarized in Table~\ref{tab:results}, \toolname identified 329 pairs of parallel methods, with 151 on Android 14, 125 on Android 15, and 53 on Android 16. The reduced count on Android 16 may indicate that the migration has reached a later stage. We also observed that parallel methods associated with \textsf{Type I (2) -- Directory Shadowing} were mostly related to multimedia functionalities, while those under \textsf{Type II} were mostly permission-related. This suggests that different development teams adopt different naming conventions and migration strategies when transitioning from Java to Kotlin. There appears to be no consistent standard across AOSP development.

From these parallel method pairs, \toolname identified 372 behavioral divergences. During the Security Assessment phase, it flagged 37 instances as potentially vulnerable. We conducted a comprehensive manual review of these 37 instances and found several duplicates. This duplication occurred for two primary reasons. First, identical vulnerabilities naturally persisted across different Android OS versions (e.g., from Android 14 to 15). Second, the LLM flagged multiple distinct behavioral anomalies within certain complex methods (e.g., a failing guard check followed by an asynchronous state mutation) that originated from the same underlying architectural flaw. After deduplicating overlapping reports, we identified 17 unique, security-relevant divergences.

We subjected the 17 unique divergences to strict manual verification through source code auditing and PoC exploit construction. We strictly adhered to our threat model to ensure the constructed attacks explicitly breached Android security boundaries. This review identified 6 false positives that did not constitute actual vulnerabilities. We categorize these false positives into two root causes. \textit{Impractical Exploitability (5 instances):} The system correctly identified genuine behavioral divergences lacking practical exploitability. The vulnerable logic was either shielded from untrusted IPC paths (lacking accessible API boundaries) or required unrealistic preconditions, such as pre-existing \texttt{root} privileges. \textit{Architectural Delegation (1 instance):} The LLM failed to strictly adhere to the filtering prompts. It misclassified a benign structural refactoring as a guard bypass because the validation logic was safely delegated to a deeper opaque method.

For the remaining 11 distinct divergences, we successfully constructed PoC exploits or verified the flaws within the AOSP source code, confirming them as genuine, exploitable vulnerabilities.

\begin{table*}[t]
\centering
\caption{Summary of Google confirmed issues.}
\label{tab:bugs}
\small \vspace{-5pt}
\begin{tabu} to \linewidth {*X[0.2,c] | X[2,l] * X[0.75,c] | X[0.9,c] | X[1.5,c] | X[1.8,l]| X[1.1,l] *}
\thickhline
\textbf{No.} & \textbf{Method} & \textbf{Issue Type} & \textbf{Affected Side} & \textbf{Affected Android OS} & \textbf{Google Confirmation} & \textbf{CVE ID}\\
\thickhline
\texttt{1} & \texttt{onPackageAdded} & EoP & Java & 14 & Vulnerability (High severity)& \texttt{CVE-2024-43095}\\

\hline
\texttt{2} &\texttt{removePermission} & EoP & Java &  14 & Vulnerability (High severity)& \texttt{CVE-2026-0026}\\
\hline

\texttt{3} &\texttt{getAppOpPermissionPackages} & InfoLeak & Kotlin &  15, 16 & Vulnerability (Low severity) & N/A\\
\hline
\texttt{4} &\texttt{checkPermission} & InfoLeak & Kotlin &  15, 16 & Acknowledged as a bug  & N/A\\
\hline
\texttt{5} &\texttt{getLegacyPermissionState} & InfoLeak & Java &  14 & Acknowledged as a bug  & N/A\\

\thickhline
\end{tabu}

\end{table*}

\paragraphNew{Google Confirmed Issues}
Following responsible disclosure practices, we reported all 11 verified vulnerabilities to the Google Android Security Team. As listed in Table~\ref{tab:bugs}, five issues received official confirmation. All five confirmed cases reside within the Android Permission subsystem. This concentration aligns with expectations. Permission management forms the most security-critical boundary in the OS. Its aggressive Kotlin migration in recent releases makes it highly susceptible to semantic translation errors.

Google classified three confirmed issues as security vulnerabilities. Two represent \textit{High-Severity Privilege Escalation (EoP)} flaws originating from the legacy Java implementations within the Android 14 path. One constitutes a \textit{Low-Severity Information Leakage} vulnerability introduced in the Kotlin implementations of Android 15 and 16. Google acknowledged the remaining two cases as functional bugs involving internal metadata exposure. The other six reported cases currently await final triage and confirmation (see Table~\ref{tab:extrabugs} in the Appendix).

\section{Case Studies}
\label{sec:bugs}

In this section, we present two representative cases to demonstrate the security implications of divergences between Java and Kotlin parallel implementations.

\subsection{Case 1: EoP via \texttt{onPackageAdded}} 
A malicious app can escalate its privileges by exploiting inconsistent permission update logic during package updates. This allows it to silently gain sensitive permissions, such as \texttt{CALL\_PHONE}.

\begin{figure}[t]
	\centering
	\includegraphics[width=1\columnwidth]{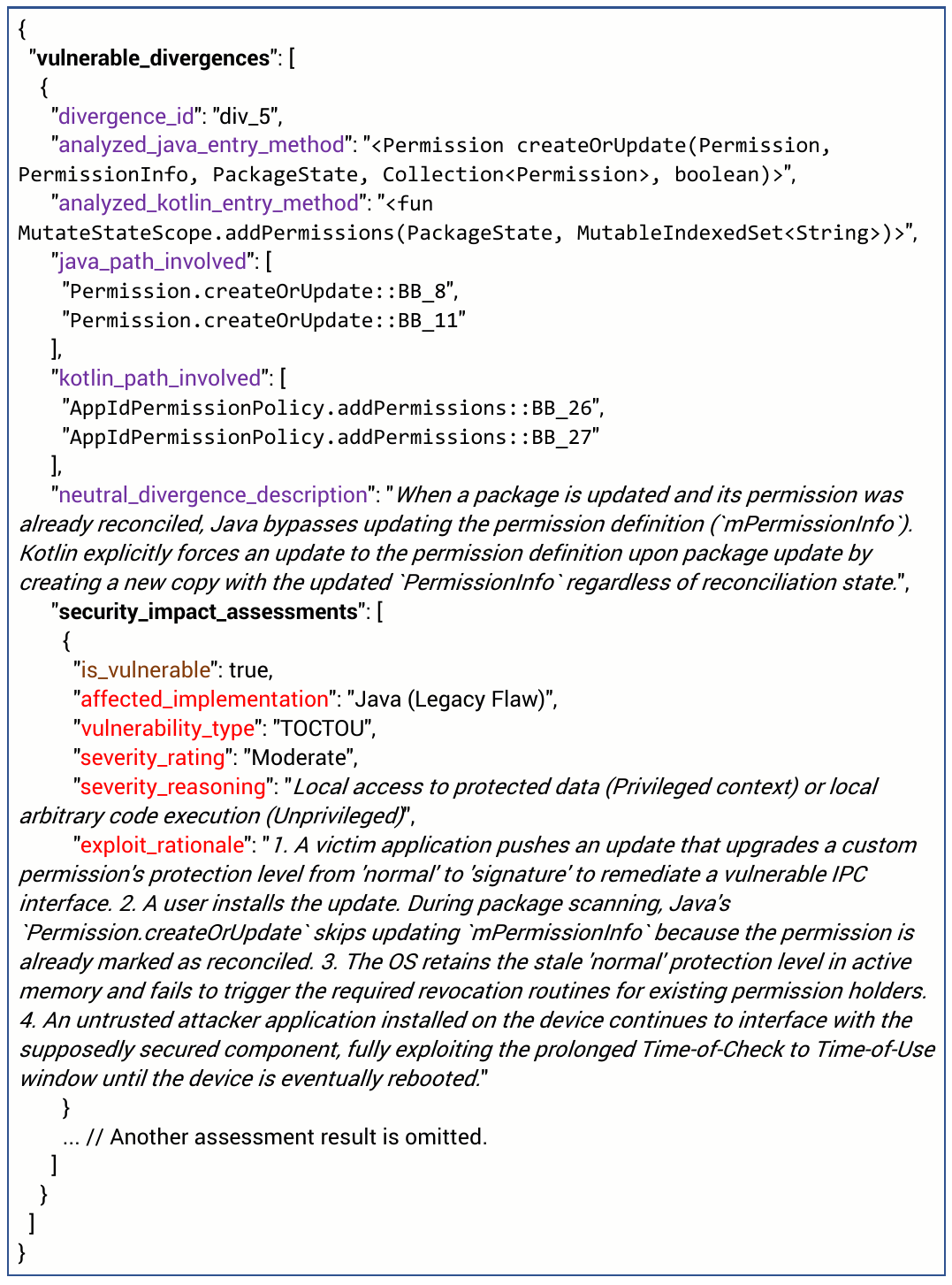}
	\caption{Report for Case 1.}
	\label{fig:case1}
\end{figure}

\paragraphNew{Behavior Divergence} 
The \texttt{onPackageAdded} method is responsible for managing adjustments to permission registration information in the system that arise from app installation or updates. As shown in Figure~\ref{fig:case1}, in the Java implementation (active in Android 14), this process relies on the \texttt{createOrUpdate} function~\cite{url_Permission_java}, which updates a permission’s registration information only if the internal flag \texttt{Permission.mReconciled} is set to \texttt{false}. Once this flag is marked as \texttt{true}, the definition remains unchanged in future updates.

In contrast, the Kotlin implementation~\cite{url_AppIdPermissionPolicy_kt} utilizes the \texttt{addPermi- ssions} function, which unconditionally updates the permission registration, ignoring the reconciliation status. Consequently, during permission updates, the Java implementation may preserve the stale (and potentially less restrictive) version, whereas the Kotlin implementation correctly applies the updated definition.

This inconsistency enables a dangerous scenario: a \texttt{normal}-level dynamic permission (declared by invoking the \texttt{addPermission} API) initially granted to an app can be silently changed to a \texttt{dangerous} one through app update while retaining its grant status. This kind of escalation would always be blocked for manifest permissions (declared in the app's manifest), where Android enforces checks to prevent protection level upgrades from \texttt{normal} to \texttt{dangerous}~\cite{DBLP:conf/sp/LiDLDG21}. However, these checks can be bypassed with dynamic permissions and the flawed update flow in the Java implementation.

\paragraphNew{Exploit} 
We developed a PoC app, \textsf{app-dynamic}, which initially registers a custom dynamic permission \texttt{com.dynamic.cp} with the \texttt{normal} protection level via \texttt{addPermission}. In a subsequent update, the app:

\begin{itemize}[leftmargin = 15pt, topsep=2pt]
    \item Updates \texttt{com.dynamic.cp} to \texttt{dangerous} and assigns it to the \texttt{PHONE} group via \texttt{addPermission}.
    \item Requests both \texttt{com.dynamic.cp} and \texttt{CALL\_PHONE}.
    \item Simultaneously declares \texttt{com.dynamic.cp} in its manifest as a \texttt{normal} permission.
\end{itemize}

On a Pixel 7a running Android 14 (Java implementation), we install \textsf{app-dynamic}, run it, and then update the app. After the update, the app automatically obtains the system permission \texttt{CALL\_PHONE} during its runtime, without prompting the user.

\paragraphNew{Root Cause} During an app update, Android revokes any permissions that are absent from the updated app manifest. Therefore, to retain the initial grant state (a prerequisite for the escalation), \textsf{app-dynamic} must declare \texttt{com.dynamic.cp} in its updated manifest. 
In the Java path, the \texttt{mReconciled} flag causes the system to ignore this new manifest declaration, because dynamic permissions are always reconciled, leaving the existing dynamic permission object untouched. This allows the subsequent dynamic update (to \texttt{dangerous}) to succeed on the existing object.
Conversely, the Kotlin path unconditionally refreshes the permission definition. This causes the manifest declaration to override the dynamic one. Since Android prohibits modifying manifest-defined permissions via \texttt{addPermission}, the subsequent escalation attempt is blocked.

\paragraphNew{Impact} 
This vulnerability allows a malicious app to silently acquire high-privilege system permissions through standard updates. The issue was confirmed by the Android Security Team and assigned \texttt{CVE-2024-43095} with a \textbf{High Severity} rating.

\subsection{Case 2: EoP via \texttt{removePermission}}
\label{case:removePermission}

In this case, we demonstrate how a discrepancy in error handling between Java and Kotlin leads to a privilege escalation vulnerability.

\begin{figure}[t]
	\centering
	\includegraphics[width=1\columnwidth]{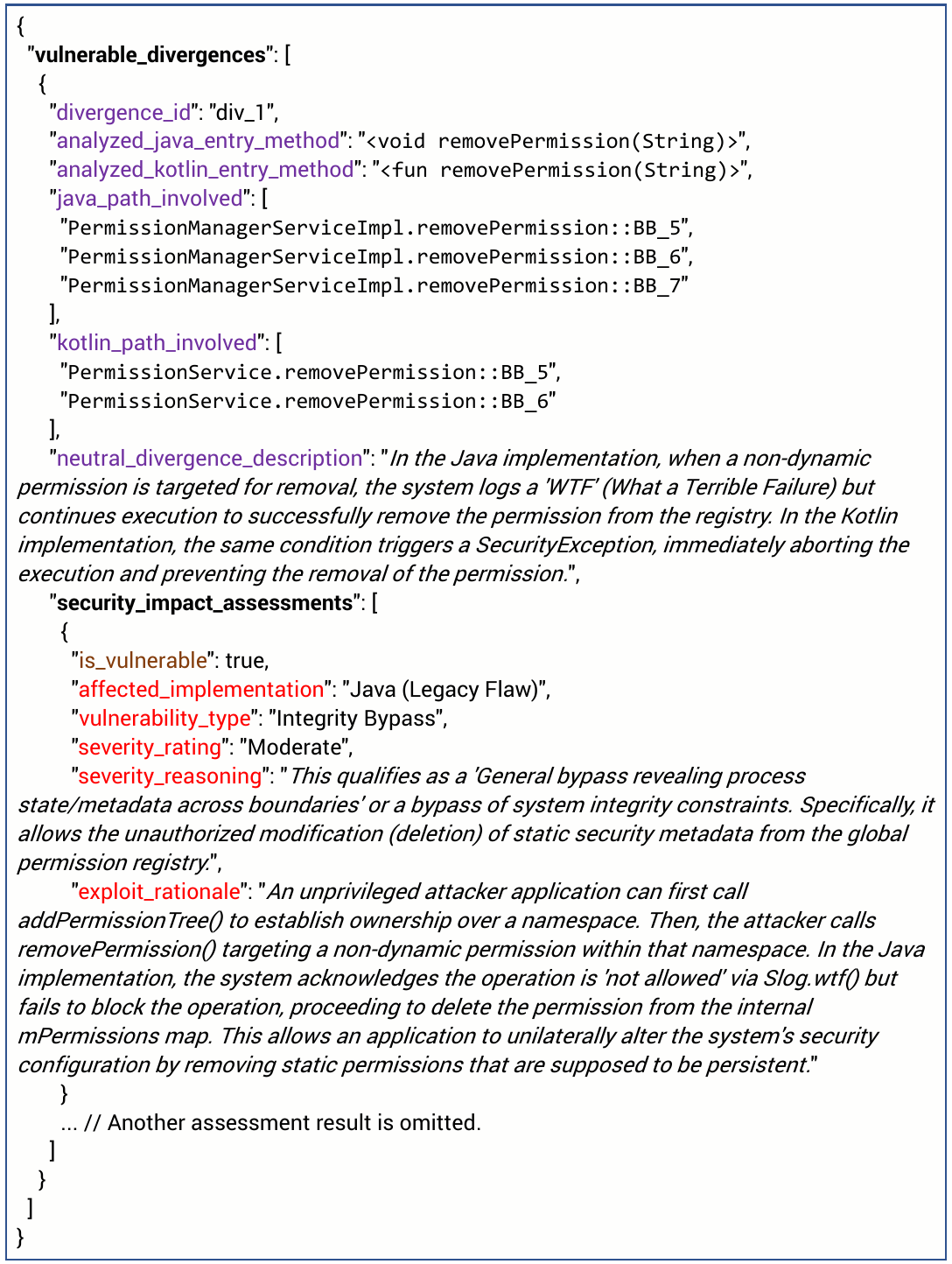}
	\caption{Report for Case 2.}
	\label{fig:case2}
\end{figure}

\paragraphNew{Behavior Divergence} 
The \texttt{removePermission} method is responsible for removing a permission definition from the system registry. 
As shown in Figure~\ref{fig:case2}, in the Java implementation (active in Android 14), the logic exhibits a fail-open behavior: it merely logs a warning (\texttt{Slog.wtf}) when encountering a non-dynamic permission but \textit{proceeds} to execute the deletion. Developers explicitly noted this temporary relaxation with a comment: ``\texttt{\textit{// TODO: switch this back to SecurityException}}''.

In contrast, the Kotlin implementation acts as a \textbf{fail-secure} guard. Under the same condition (attempting to remove a non-dynamic permission), it throws a \texttt{SecurityException}, effectively aborting the operation.

\paragraphNew{Exploit} 
This behavioral divergence implies that systems running the Java implementation are vulnerable to unauthorized permission deletion, whereas the Kotlin path blocks such attempts. 
We designed a PoC attack on a Pixel 7a running Android 14. The exploitation proceeds in two steps:

\miniparagraph
(1) \textit{Permission Deletion:} We developed an \textsf{attacking-app} that invokes \texttt{removePermission} to delete the system-defined \texttt{READ\_VOICEMAIL} permission. Due to the Java implementation's fail-open error handling, the permission is successfully removed from the registry.

\miniparagraph
(2) \textit{Ownership Takeover:} The app subsequently updates itself to declare \texttt{READ\_VOICEMAIL} as a custom permission with a \texttt{normal} protection level (see Listing~\ref{code:attacking-app}).

\begin{lstlisting}[language=Xml, 
	style=mystyle,
    literate={@}{{@}}1,
    mathescape=false,
	basicstyle=\ttfamily\footnotesize, 
	caption={Manifest snippet of \textsf{attacking-app}.}, 
	label={code:attacking-app}]
<!-- Step 2: Declare the previously system-defined permission as normal -->
<permission android:name="com.android.voicemail.permission.READ_VOICEMAIL"
android:protectionLevel="normal">
</permission>
<uses-permission android:name="com.android.voicemail.permission.READ_VOICEMAIL">
</uses-permission>
\end{lstlisting}

Our experiment confirmed that the \textsf{attacking-app} successfully obtained ownership of the \texttt{READ\_VOICEMAIL} permission. Consequently, the app accessed the user's private system voicemail, bypassing the intended access control. This constitutes a privilege escalation vulnerability. 

\paragraphNew{Impact} 
We responsibly disclosed this vulnerability to the Android Security Team. The issue was confirmed and assigned \texttt{CVE-2026-0026} with a \textbf{High Severity} rating. This case reveals that legacy Java paths, if left active and inconsistent with stricter Kotlin implementations, can serve as latent attack vectors.

\section{Discussion}
\label{sec:discussion}

We next discuss broader implications, current limitations, and practical mitigation strategies.

\paragraphNew{Broader Implications: Migration Security} Beyond the specific context of Android, our findings suggest potential security implications for broader software evolution. As the industry increasingly adopts modern languages, such as migrating Linux kernel subsystems to Rust or legacy Objective-C iOS components to Swift, legacy and modern implementations often coexist to support staged rollouts. This coexistence creates a distinct attack surface where subtle semantic gaps between languages, such as in error propagation or type safety, can silently compromise security invariants. Therefore, this work underscores the importance of \textit{Migration Security}. We position automated differential analysis not merely as a bug-finding technique but as a promising assurance mechanism for the hybrid future of critical software infrastructure.

\paragraphNew{LLM-assisted Analysis} Our approach employs LLMs to evaluate behavioral equivalence across Java and Kotlin. To minimize ambiguity, we provide comprehensive code context, including call chains and method dependencies. To address the probabilistic nature of generative models, we enhance stability through structural grounding, constrained CoT, contextual verification, RAG, and a majority-voting strategy. Although LLMs occasionally produce non-deterministic outputs, combining rich context with consensus mechanisms filters out transient inconsistencies. Consequently, potential inaccuracies primarily manifest as conservative false positives. We mitigate these remaining inaccuracies through manual verification.

\paragraphNew{Mitigation Strategies for Parallel Implementations} 
Our findings suggest actionable strategies to reduce security risks during language migration. For platform vendors and system developers, we recommend enforcing a strict migration policy that designates a single source of truth for each subsystem. Once the new implementation achieves functional parity, the legacy version should be deprecated and removed immediately. During the transition period, behavioral equivalence should be verified using automated differential testing. This involves executing both versions on identical inputs and strictly comparing their outputs, side effects, and security checks. Additionally, code intended for future versions should be excluded from current production builds to minimize the attack surface. For end users, since these vulnerabilities reside within the system framework, the primary mitigation is to keep devices updated and apply security patches promptly.

\section{Related Work}
\label{sec:relatedwork}

Android security has long been an active research area, and many studies have been conducted~\cite{DBLP:conf/uss/LiuZGZRZ20, DBLP:conf/sp/LiDLDG21, DBLP:conf/uss/JiE0S21, DBLP:conf/uss/LiDYLGZ23,DBLP:conf/uss/GorskiTEC22,DBLP:conf/ccs/XiangZWGL21,DBLP:conf/uss/MaarDLM24,DBLP:conf/ndss/AcarTLOAU24}. We focus our review on related work in Android differential analysis and cross-language security analysis.

\paragraphNew{Differential Analysis} Differential analysis is an effective approach for security analysis. It has been widely used to uncover security flaws introduced by OEM customization, ecosystem fragmentation, and software evolution by comparing a customized or updated target with a trusted baseline. Bandara et al.~\cite{DBLP:conf/eurosp/BandaraPGKVTV25} performed large-scale differential analysis of OEM-customized Android TLS stacks against the AOSP baseline and showed that security-critical verification logic is frequently modified in ways that can weaken app-level TLS security. Dai et al.~\cite{10.1145/3763141} proposed ApkDiffer, a two-stage decomposition-based diffing tool that aligns functionally equivalent methods across app versions to reduce alignment errors and enable precise security-relevant change localization. Continella et al.~\cite{DBLP:conf/ndss/ContinellaFLPZK17} made black-box differential analysis practical for Android privacy leak detection by controlling sensitive inputs and eliminating network nondeterminism to infer leaks from traffic deviations even under obfuscation. Aafer et al.~\cite{DBLP:conf/uss/AaferZD16} systematically identified customization-sensitive security features and used large-scale differential analysis across 591 custom images to find prevalent inconsistencies that they validated as exploitable on real devices. Zhou et al.~\cite{DBLP:conf/sp/ZhouLZNW14} built ADDICTED to differentially compare device file protections between customized phones and official Android, exposing under-protected driver interfaces that enable unprivileged app attacks. Yang et al.~\cite{DBLP:conf/issre/YangBLGD24} performed differential analysis between Google Play and third-party market versions of apps claiming the same version code, uncovering widespread security and privacy inconsistencies.

Unlike these studies, our work does not perform differential analysis in the conventional setting, which typically assumes a single language or the same target type. Instead, we focus on Java and Kotlin, which introduce a series of challenges from pair identification to security analysis.

\paragraphNew{Cross-language Analysis} The development of Android apps and system components can involve multiple programming languages, including Java, Kotlin, C/C++ (via the NDK), HTML, and JavaScript. Recent work has explored cross-language analysis in this context. For mixed analysis of Java and web programming languages, Hu et al.~\cite{DBLP:conf/issta/HuW0C23} developed \(\omega\)Test, a test generation technique for Android WebViews. It applies cross-language dynamic analysis of Java and JavaScript to capture WebView-specific properties and generate event sequences for detecting interaction-related bugs. Tiwari et al.~\cite{DBLP:conf/issre/TiwariPH23} designed a demand-driven analysis framework for Android hybrid apps, which tracks information flows between Java code in the app and embedded JavaScript by selectively summarizing the shared Java code based on its usage in JavaScript.

For mixed analysis of Java code and native code (C/C++) in Android apps, Xiong et al.~\cite{DBLP:conf/issta/XiongDCQWSZ24} proposed Atlas, a cross-language fuzzing framework for Android closed-source native libraries. It analyzes both Java and native code to generate harnesses and supports fuzzing in an emulator with a Java runtime. Wang et al.~\cite{DBLP:conf/issta/Wang024} introduced NativeSummary, an inter-language static analysis framework for Android apps. It extracts semantics from native C/C++ code, translates JNI usage and native calls into Java bytecode, and extends existing Java analysis tools to support cross-language data flow analysis. Borzacchiello et al.~\cite{DBLP:conf/esorics/BorzacchielloCM22} developed DroidReach, a static analysis approach that combines heuristics and symbolic execution to accurately assess the reachability of native C/C++ functions in Android apps, thereby enabling more precise vulnerability assessment.

Unlike prior studies that target individual Android applications, our work analyzes system-level Android OS code. Consequently, the identified security issues affect AOSP and broadly impact downstream device vendors.

\section{Conclusion}
\label{sec:conclusion}

To our knowledge, this paper presents the first systematic security investigation of parallel Java and Kotlin implementations within the Android framework. We designed \toolname to automatically discover parallel method pairs and perform cross-language differential analysis. By combining bytecode-level IR, a language-agnostic Unified Execution Graph (UEG), and an LLM-assisted reasoning engine, \toolname effectively bridges the semantic gap to identify behavioral deviations. Evaluated on AOSP Android~14--16, \toolname identified 372 divergences across 329 method pairs, flagging 37 as potentially vulnerable. Responsible disclosure yielded three confirmed vulnerabilities (with two CVE IDs) and two functional bugs. Ultimately, our findings highlight a critical takeaway: OS-level language migrations are not merely syntax updates, but inherently security-relevant transformations that require systematic semantic consistency checks.

\appendix

\section*{Ethical Considerations}

\paragraphNew{Experimental Safety} All analyses in this study were conducted in a controlled, isolated environment without interacting with live user devices or external cloud services. This design choice ensured that no end users, applications, or third-party services were exposed to unintended risks or service disruptions during our evaluation.

\paragraphNew{Responsible Disclosure} We have followed a coordinated vulnerability disclosure process with the Android Security Team. To date, five reports have been confirmed: three as vulnerabilities and two as bugs. Two CVE IDs have been assigned: \texttt{CVE-2024-43095} (high severity) and \texttt{CVE-2026-0026} (high severity). Google acknowledged our findings by awarding us USD 14,000 through the Android Security Reward Program.

\paragraphNew{Ecosystem Impact Management} We acknowledge that the identified vulnerabilities reside in the core AOSP framework and thus affect the broader Android ecosystem, including downstream vendors (e.g., Samsung, Xiaomi, OPPO). To mitigate the risk of potential exploitation before patches propagate to customized system images, we have strictly adhered to the disclosure timeline and withheld proof-of-concept exploits from the public domain until adequate mitigation measures are available.

\paragraphNew{Research Purpose} We believe that responsibly sharing these findings benefits the community by providing platform maintainers and researchers with the necessary insights to address this emerging class of cross-language security risks.

\balance
\bibliographystyle{plain}
{\normalem
\bibliography{refs}
}

\appendix

\section{LLM Selection Experiment}
\label{appendix:llm-select}

To select the most appropriate Large Language Model (LLM) for our highly specific code analysis tasks, we conducted an empirical comparison using the ground-truth dataset established in Section~\ref{subsec:ablation}. Specifically, we evaluated the performance of three state-of-the-art models: \textsf{Gemini 3.1 Pro}, \textsf{GPT-5.4}, and \textsf{Claude Opus 4.7}. 

All candidate models were integrated into the \toolname pipeline and tasked with processing the same Unified Execution Graphs (UEGs) under identical prompt constraints. We specifically restricted our comparative evaluation to the \textit{Divergence Identification} stage. The rationale is straightforward: accurate divergence extraction is a strict prerequisite for downstream threat modeling. If an LLM generates excessive structural noise or fails to properly align the graphs at this foundational stage, its subsequent security assessments become inherently unreliable. Therefore, we measured the \textit{Coverage} (the percentage of genuine ground-truth divergences identified) and \textit{Accuracy} (the percentage of reported divergences that are actually genuine).

The experimental results are detailed in Table~\ref{tab:llm-selection}. While \textsf{Claude Opus 4.7} achieved the highest coverage (92.2\%), it exhibited a remarkably low accuracy (59.5\%), suggesting it aggressively hallucinated non-existent semantic divergences. \textsf{GPT-5.4} performed poorly across both metrics. \textsf{Gemini 3.1 Pro} provided the optimal balance, achieving a highly competitive coverage (80.4\%) while decisively leading in accuracy (87.2\%). Because minimizing false positives early in the analysis pipeline is critical to preventing alert fatigue and cascading errors during the threat modeling phase, \textsf{Gemini 3.1 Pro} was selected as the default inference engine for \toolname.

\section{The Supplementary Results}
\label{appendix:supplementary}

As discussed in Section~\ref{subsec:results}, manual inspection and deduplication identified 11 vulnerable divergences. Table~\ref{tab:bugs} details the issues officially confirmed by the Android Security Team. Table~\ref{tab:extrabugs} catalogs the remaining issues currently under investigation or pending final resolution. This inclusion provides a comprehensive view of the discovery scale achieved by \toolname.

\begin{table}[h]
	\centering
	\caption{Performance comparison of candidate LLMs.}
	\label{tab:llm-selection}
	\small  \vspace{-5pt}
	\begin{tabu} to \linewidth {* X[1.5,l] | X[2,c] | X[2,c] *}
		\thickhline
		\textbf{Model} & \textbf{Coverage} & \textbf{Accuracy} \\
		\thickhline
		\textsf{Gemini 3.1 Pro} & 80.4\% & 87.2\% \\ 
		\hline
		\textsf{GPT-5.4} & 70.6\% & 41.9\% \\ 
		\hline
		\textsf{Claude Opus 4.7} & 92.2\% & 59.5\% \\ 
		\thickhline
	\end{tabu}
\end{table}

\begin{wraptable}{r}{0.474\textwidth}
	% \begin{table}[t]
		\centering
		\caption{Summary of cases under investigation.}
		\label{tab:extrabugs}
		\small \vspace{-5pt}
		\begin{tabu} to \linewidth {* X[0.3,c] | X[2.6,l] | X[1,c] | X[1,c] | X[1.5,c] *}
			\thickhline
			\textbf{No.} & \textbf{Method} & \textbf{Issue\newline Type} & \textbf{Affected\newline Side} & \textbf{Affected\newline Android OS} \\
			\thickhline
			6 & \texttt{getAllowlistedRes-}\newline \texttt{trictedPermissions} & InfoLeak & Kotlin & 15, 16 \\
			\hline
			7 & \texttt{getAllPermission- Groups} & InfoLeak & Java & 14 \\
			\hline
			8 & \texttt{isPermissionRevoked-}\newline \texttt{ByPolicy} & InfoLeak & Kotlin & 15, 16 \\
			\hline
			9 & \texttt{onPackageAdded} & EoP & Java &  14 \\
			\hline
			10 & \texttt{onPackageAdded} & EoP & Kotlin &  15, 16 \\
			\hline
			11 & \texttt{systemReady} & EoP & Kotlin &  15, 16 \\
			\thickhline
		\end{tabu}
		% \end{table}
\end{wraptable}
\balance

\end{document}